# Generation of wave packets and breathers by oscillating kinks in the sine-Gordon system

M.M. Bogdan[1,2], O.V. Charkina[1]

[1]*B. Verkin Institute for Low Temperature Physics and Engineering of the National Academy of Sciences of Ukraine,
47 Nauky Ave., Kharkiv, 61103, Ukraine*
[2]*V.N. Karazin Kharkiv National University, 4 Svobody Sq., Kharkiv, 610022, Ukraine*

E-mail: m_m_bogdan@ukr.net
charkina@ilt.kharkov.ua

**Abstract**

Evolution of the nonequilibrium inhomogeneities and topological defects is studied in terms of complex kink solutions of the sine-Gordon equation. The weakly damped oscillation of the sine-Gordon kink, named as the kink quasimode, is described explicitly. It is shown that the oscillatory kink behavior and the wave packet generation depend significantly on the initial nonequilibrium kink profile. In order to specify conditions of the generation of wobbling kinks with a multibreather structure we reformulate the direct scattering problem associated with the SG equation as the spectral problem of the Schrödinger operator. We obtain the dependence of the radiation energy, which is emitted during formation of the multi-frequency wobbling kink, on the effective dimension of its initial profile.

**Introduction**

The structure and dynamics of topological defects and inhomogeneities in solid state physics are described by nonlinear equations. Well-known examples of such topological objects are domain walls in anisotropic magnets, fluxons in long Josephson junctions, and dislocations in crystals. In general, locally inhomogeneous fields created by these objects look like kink configurations when two parts of the system lie in neighboring equivalent potential minima separated by energy barrier. Corresponding field variables are described by the kinks, one-parametric soliton solutions of the nonlinear equations.

The sine-Gordon (SG) equation and its kink solution are most famous [1] due to the complete integrability of the SG model. This model describes in the explicit form the stationary dynamics of magnetic domain walls, fluxons, and dislocations mentioned above. In the framework of the SG equation, the kink as a true soliton moves freely with a constant velocity and an unchanged profile. Topological solitons can be excited dynamically through the formation of soliton-antisoliton pair as a result of the decay of a large-amplitude breather, that is considered as the soliton-antisoliton oscillating bound state. On the other hand, introducing a single topological kink in a medium can be done only through a boundary or interface. In this case, it would be expected that an initial kink profile does not coincide with stationary one. Additionally, a variety of physical factors disturbing the integrability of the SG model could affect the particle-like behavior of the kink. In particular, the kink shape can stop to be a rigid structure. Then besides a static deformation, the shape becomes able to oscillate around a previously stationary profile under the action of both kinds of perturbations, either nonequilibrium initial conditions or disturbing the equation. As a result, a generation of the oscillating kink appears to be possible. If the amplitude of oscillations is small and localized in space, it means an existence of the internal mode in the linear excitation spectrum of the kink [2]. If the amplitude is not small, then the nonlinearity of the SG equation becomes essential, and a complex oscillating solution can be considered as a combination of the kink and the breather, which is a time-periodic and space-localized nonlinear excitation of the system. Such an exact combined solution of the SG equation is well-known and called the wobbling kink or the wobble [3,4].

Although a mathematical theory of solitons establishes strictly a full set of excitations of the SG model, namely kinks, antikinks and breathers, and linear waves of continuous spectrum, the existence of the internal oscillating mode and the quasimode in the SG system, as collective excitations of the kink, is discussed for a long time [5,6]. Usually, reports about theoretical revealing and further observation of these modes are based on variational or asymptotical approaches to the problem [7,8] and numerical simulations of the soliton dynamics under perturbation of the SG equation or initial conditions for the kink [9,10]. Then every finding of quasi- and internal modes forces to seek for physical reasons and factors of their appearance in numerical experiments.

In general, the internal modes of kinks are responsible for nontrivial inelastic interaction of kinks in nonintegrable nonlinear equations, as it takes place in the $\varphi^4$- model and the double sine-Gordon system [11,12]. Corresponding equations possess topological soliton solutions, a kink and a kink-kink bound state called a wobbler, respectively, with well-defined internal modes. Discrete solitons also can possess such kind of localized modes. The sine-Gordon equation is derived from the discrete sine-Gordon equation describing the Frenkel-Kontorova model for dislocations [13]:

$$\frac{\partial^2 u_n}{\partial t^2} + 2u_n - u_{n-1} - u_{n+1} + \frac{1}{d^2}\sin u_n = 0 \ . \qquad (1)$$

Depending on applications, the function $u_n$ can denote either atomic displacements in a crystal as in the Frenkel-Kontorova model or the double azimuthal angle of rotation of spins in the easy plane in the discrete model of an anisotropic magnetic chain, as well as the phase difference of the wave functions of superconductors in a discrete set of the Josephson junctions. The parameter $d$ is a measure of the discreteness of the model. The integrable SG equation is the long-wavelength limit of the discrete equation (1):

$$u_{tt} - u_{xx} + \sin u = 0 .\qquad(2)$$

Taking into account a higher dispersion term in the expansion of the second difference $u_{n-1} + u_{n+1} - 2u_n \approx u_{xx} + \beta u_{xxxx}$, where $\beta = 1/12d^2$ is a dispersive parameter, we obtain the dispersive SG equation. To void artificial instability of linear excitations, we perform the transition from the equation with fourth spatial derivatives to regularized sine-Gordon equation (RSGE) [14-18]:

$$u_{tt} - u_{xx} - \beta u_{xxtt} + \sin u = 0 .\qquad(3)$$

The discrete kink of the equation (1) has the internal mode with an extremely weak localization [19], and in the case of a large enough parameter $d$ and respectively a small $\beta$ this feature of the kink behavior can be reproduced in the framework of the dispersive SG equations [15,20]. Moreover, we showed [2,21] that the static kink of the RSGE, having the same form as a kink of the SG equation

$$u_K = 4\arctan\exp(x) ,\qquad(4)$$

can possess a whole set of internal modes, depending on the value of the parameter $\beta$. If $\beta$ tends to unity, then a continuous spectrum of the kink excitations degenerates, and a number of internal modes become infinite. Thus a character of the kink dynamics and interaction of kinks in the RSGE depends on the strength of dispersion. In particular, in the case of small dispersion, we studied analytically a nonstationary motion of a kink having an initial profile of the SG equation kink in the framework of the RSGE and described explicitly consistent oscillatory behavior of its effective width and velocity [14]. The influence of a true internal mode appears to be negligible in this case, while the quasimode plays a principal role.

The effect of nonlinearity on developing the internal mode is traditionally studied by use of different kinds of mathematical procedures [22,23] constructing formal asymptotic expansions. However, these approaches usually have a restricted application area as it was pointed out still by Segur in Ref. [3]. At the same time, it appears that nonlinearity itself can produce a mechanism of generation of internal kink oscillation, which develops into a breather settled on the kink.

In the present work, we described explicitly the oscillatory regime of the quasimode of the nonequilibrium SG kink, introduced in Ref. [5]. Its frequency lies in the continuous spectrum and tends to the lowest frequency edge from the above. It is shown that the oscillatory behavior of the quasimode, in particular the frequency approach to the lowest frequency, depends significantly on the initial nonequilibrium kink profile. In the framework of the SG equation, we describe exactly the conditions of the generation of wobbling kinks with a multibreather structure as a result of the evolution of the nonequilibrium SG kink profile. For this purpose, we analyze the direct scattering problem associated with the SG equation and find a dependence of the radiation energy on the effective dimension of the initial kink profile.

## Quasi-local oscillations of kinks in the sine-Gordon equations

Earlier, we theoretically studied the internal oscillations of a static kink of the regularized equation (3) [2,15,21]. We also found analytically and numerically a number of features of its nonstationary motion and show efficiency of a simple analytical approach consisting in the use of the perturbation theory in the case of the weak dispersion, i.e., for a small value of the parameter $\beta$. In particular, the problem of evolution of the initial SG kink in the RSGE was solved and shortly reported in Ref. 14. Here we show that the found solutions can be used to solve the quasimode problem about oscillation of the perturbed SG kink profile in the SG equation (2).

In Ref. 14, in the framework of the RSGE, we studied the motion of the SG kink $u_K(z)$ obtained from (4) by the Lorentz transformation of coordinates $z = (x - Vt)/\sqrt{1-V^2}$, where $V$ is the initial value of kink velocity. We sought a nonstationary solution to equation (3) in the form:

$$u(x,t) = u_K(z) + u_1(z,\tau), \tag{5}$$

where $\tau = (t - Vx)/\sqrt{1-V^2}$ and the small additional function to the kink obeys the linearized equation:

$$\left(\frac{\partial^2}{\partial \tau^2} + L\right)u_1 \equiv u_{1\tau\tau} - u_{1zz} + \left(1 - \frac{2}{\cosh^2 z}\right)u_1 = \beta(u_{2\pi}(z) - u_1)_{xxtt}. \tag{6}$$

Because of the small value of $\beta$ and smallness of the additional function with respect to $u_K(z)$, we neglected $u_1$ on the right side of Eq. (6). Thus we did not take into the very weakly localized internal mode of the RSGE kink [15,21]. As a result of evolution, the RSGE kink tends to a quasi-equilibrium state with the static addition to its profile:

$$\Delta u(z) = \alpha\left(3\frac{\sinh z}{\cosh^2 z} - \frac{z}{\cosh z}\right), \qquad \alpha = \frac{\beta V^2}{(1-V^2)^2}. \tag{7}$$

The general solution can be written as $u_1(z,\tau) = \Delta u(z) + v(z,\tau)$, where $v(z,\tau)$ is the solution to the homogeneous side of Eq. (6) (without the right hand side). Indeed, suppose that at the initial moment $u(z,0) = u_K(z)$ and $u_\tau(z,0) = 0$. This means that $v(z,0) = -\Delta u(z)$ and $v_\tau(z,0) = 0$. Using the continuous spectrum eigenfunctions of the operator $L$ [24]:

$$\psi_k(z) = \frac{1}{\sqrt{2\pi}\omega(k)}(\tanh z - ik)\exp(ikz), \qquad \omega(k) = \sqrt{1+k^2}, \tag{8}$$

we solved the initial problem for the function $v(z,\tau)$:

$$v(z,\tau) = -\frac{\alpha}{4}\int_{-\infty}^{\infty}\frac{1+3k^2}{1+k^2}\frac{\cos\left(\sqrt{1+k^2}\,\tau\right)}{\cosh\frac{\pi}{2}k}(\tanh z\cos kz + k\sin kz)dk. \qquad (9)$$

The central part of this additional function, localized on the RSGE kink, can be interpreted as oscillations of its effective dimension. The rest part of the function corresponds to the spreading radiation of the continuous spectrum waves that are responsible for the decay of the kink oscillation. As the measure of the effective length of the kink profile we used the quantity $l(\tau)$ introduced as follows:

$$\frac{1}{l(\tau)} \equiv \kappa(\tau) = \frac{1}{2}\left.\frac{\partial u(z,\tau)}{\partial z}\right|_{z=0} \qquad (10)$$

In further we call the function $\kappa(\tau)$ as the reverse kink length. In particular, for the static kink solution (4) this parameter is constant and equal to unity. For the quasi-stationary kink profile with the additional function (7) the reverse kink length is $\kappa_0 = 1+\alpha$.

In Ref. 14 the time-dependent additional function $\Delta\kappa(\tau)$ for the reverse kink length as a characteristic of the nonstationary evolution of the RSGE kink was found from equation (9):

$$\Delta\kappa(\tau) = \frac{1}{2}\left.\frac{\partial v(z,\tau)}{\partial z}\right|_{z=0} = -\frac{\alpha}{4}\int_0^\infty \frac{(1+3k^2)\cos\left(\sqrt{1+k^2}\,\tau\right)}{\cosh\frac{\pi}{2}k}dk. \qquad (11)$$

As a result, we showed that the reverse kink length could be regarded as a collective variable that describes the quasi-local oscillations of the kink with the frequency in the continuous spectrum.

It is easy to see, that the solved problem turns out to be closely related to the problem of the quasimode (quasi-local oscillation) in the integrable SG equation [5–7], and its solution is directly associated with the formulas presented above.

Indeed, in Ref. [5], the solution of the SG equation (1) was numerically sought, assuming that at the initial moment, the kink is at rest and has the nonequilibrium profile with the reverse kink length $\kappa = 1+\eta$ where $\eta \ll 1$:

$$u(x,0) = 4\arctan\exp(\kappa x) \approx 4\arctan\exp(x) + \eta\frac{2x}{\cosh x}. \qquad (12)$$

Using a smallness of the parameter $\eta$, we shall look for a solution to the SG equation in the form:

$$u(x,t) = u_K(x) + \varphi(x,t), \qquad (13)$$
$$u_t(x,0) = 0,$$

We assume that $\varphi(x,0) = 2\eta x/\cosh x$ and $\varphi_t(x,0) = 0$, and then the small addition function $\varphi$ to the kink obeys the linearized equation:

$$\left(\frac{\partial^2}{\partial t^2}+L\right)\varphi \equiv \frac{\partial^2\varphi}{\partial t^2}+\left(-\frac{\partial^2}{\partial x^2}+1-\frac{2}{\cosh^2 x}\right)\varphi=0. \tag{14}$$

Representing $\varphi(x,t)$ in the form $\int_{-\infty}^{\infty}C_k\psi_k(x)\cos(\omega(k)t)dk$ by the use of eigenfunctions (8) we find coefficients $C_k$ from the initial conditions:

$$C_k=\sqrt{\frac{\pi}{2}}\frac{1}{\omega(k)\cosh\frac{\pi}{2}k} \tag{15}$$

As a result, we obtain the following solution of the quasimode problem of the SG equation:

$$\varphi_1(x,t)=\eta\int_{-\infty}^{\infty}\frac{\cos\left(\sqrt{1+k^2}\,t\right)}{(1+k^2)\cosh\frac{\pi}{2}k}(\tanh z\cos kz+k\sin kz)dk. \tag{16}$$

Notice that the function $\varphi(x,0)$ differs from the second term in the expression (7) for the addition $\Delta u(z)$ only by a numerical factor. The evolution of this kind of the perturbation of the kink is shown in Fig. 1. Here and further we show only a half of picture on the positive axis $x$ because the left part is antisymmetric reflection of the right one due to the oddness of functions. The time interval between configurations is chosen as $\Delta t=2\pi$. As can be seen, the kink slope changes periodically and causes the radiation of extending waves. We call this oscillating kink behavior as the quasimode regime. In order to study the time behavior of a central part of the kink we again relate the addition to the reverse kink length with a half of the derivative of the function $\varphi_1(x,t)$ in the zero point and calculate numerically the corresponding integral:

$$\Delta\kappa_1(t)=\frac{1}{2}\frac{\partial\varphi_1(x,t)}{\partial x}\bigg|_{x=0}=\eta\int_0^{\infty}\frac{\cos\left(\sqrt{1+k^2}\,t\right)}{\cosh\frac{\pi}{2}k}dk. \tag{17}$$

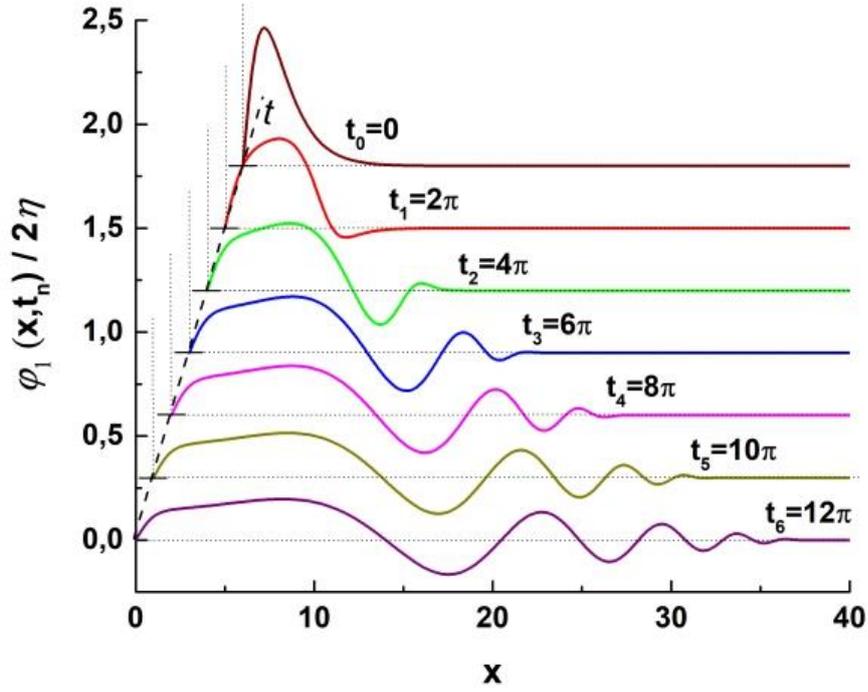

*Fig.1* The evolution of the addition to the SG kink in the quasimode regime.

Its time dependence is shown in Fig. 2. The fast Fourier transform of this time series restricted by a first hundred periods reveals a peak at frequency $\Omega_1 = 1.0048$, which is above the lowest frequency edge $\Omega_0 = 1$, and this result is in full agreement with the numerical result $\Omega_{BW} = 1.004 \pm 0.001$ by Boesch and Willis [5] and the fact that the kink has no internal mode in the SG equation in the linear approximation [2]. Analysis of the dependence (17) shows that the frequency of damped oscillations of the reverse kink length rapidly approaches the edge of the continuous spectrum from the above. Indeed the expression for $\Delta\kappa_1(t)$ can be represented in the form of $\Delta\kappa_1(t) = \eta A(t)\cos(t + \gamma(t))$, where the amplitude is equal to $A(t) = |J(t)|$ and the phase $\gamma(t) = \mathrm{Arg}\, J(t)$, and the integral $J(t)$ is defined as

$$J(t) = \int_0^\infty \frac{\exp\left(i\left(\sqrt{1+k^2}-1\right)t\right)}{\cosh\frac{\pi}{2}k}\,dk. \tag{18}$$

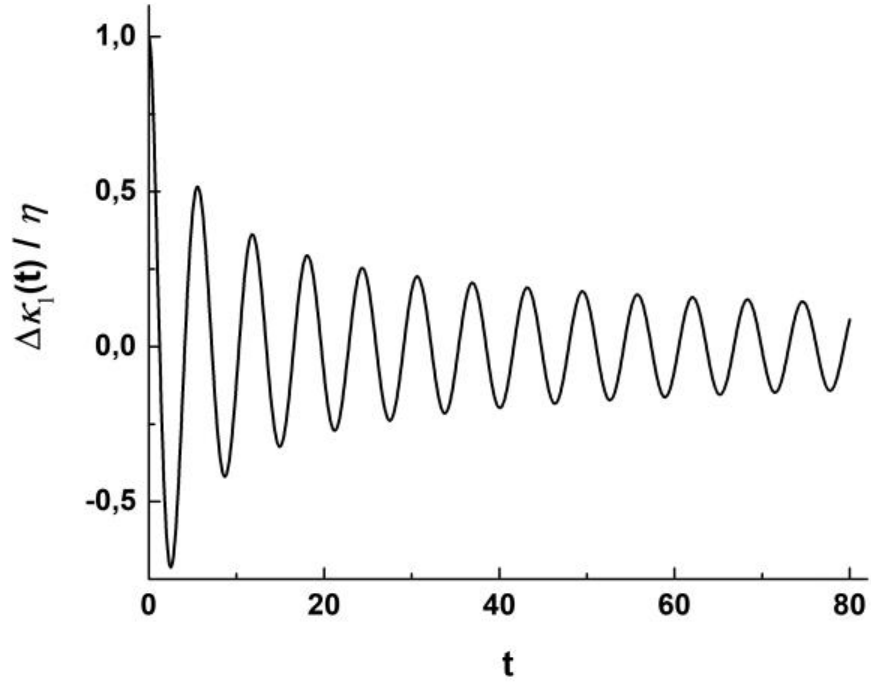

*Fig.2* Weakly damped oscillations of the addition to the reverse kink length in the quasimode regime.

The amplitude $A(t)$ is appeared to be a monotonically decreasing function, as shown in Fig. 3a. This dependence quickly approaches asymptotics $\sigma/\sqrt{t}$ with the numerical value $\sigma = 2.514$ but then slowly decreases according to this decay law. In order to estimate a contribution of phase $\gamma(t)$ in the time behavior of the reverse kink length, we construct the time derivative of the full phase $\Omega_\gamma(t) = 1 + \dfrac{d}{dt}\gamma(t)$ and derive for it the following expression

$$\Omega_\gamma(t) = 1 + \mathrm{Im}\left(\frac{d}{dt}\ln J(t)\right). \tag{19}$$

As shown in Fig. 3b, $\Omega_\gamma(t)$ is also a monotonically decreasing function, which rapidly approaches $\Omega_0 = 1$, but then it stays above at a very close distance for a long time. Thus the time dependence of the effective kink length $l(t) = 1/\kappa(t)$ indicates the quasi-local character of this type of internal oscillations, and being a collective variable $l(t)$ describes a coherent motion of particles forming the kink central part. The oscillations dissipate in virtue of a loss of the kink energy through the excitation of spreading waves.

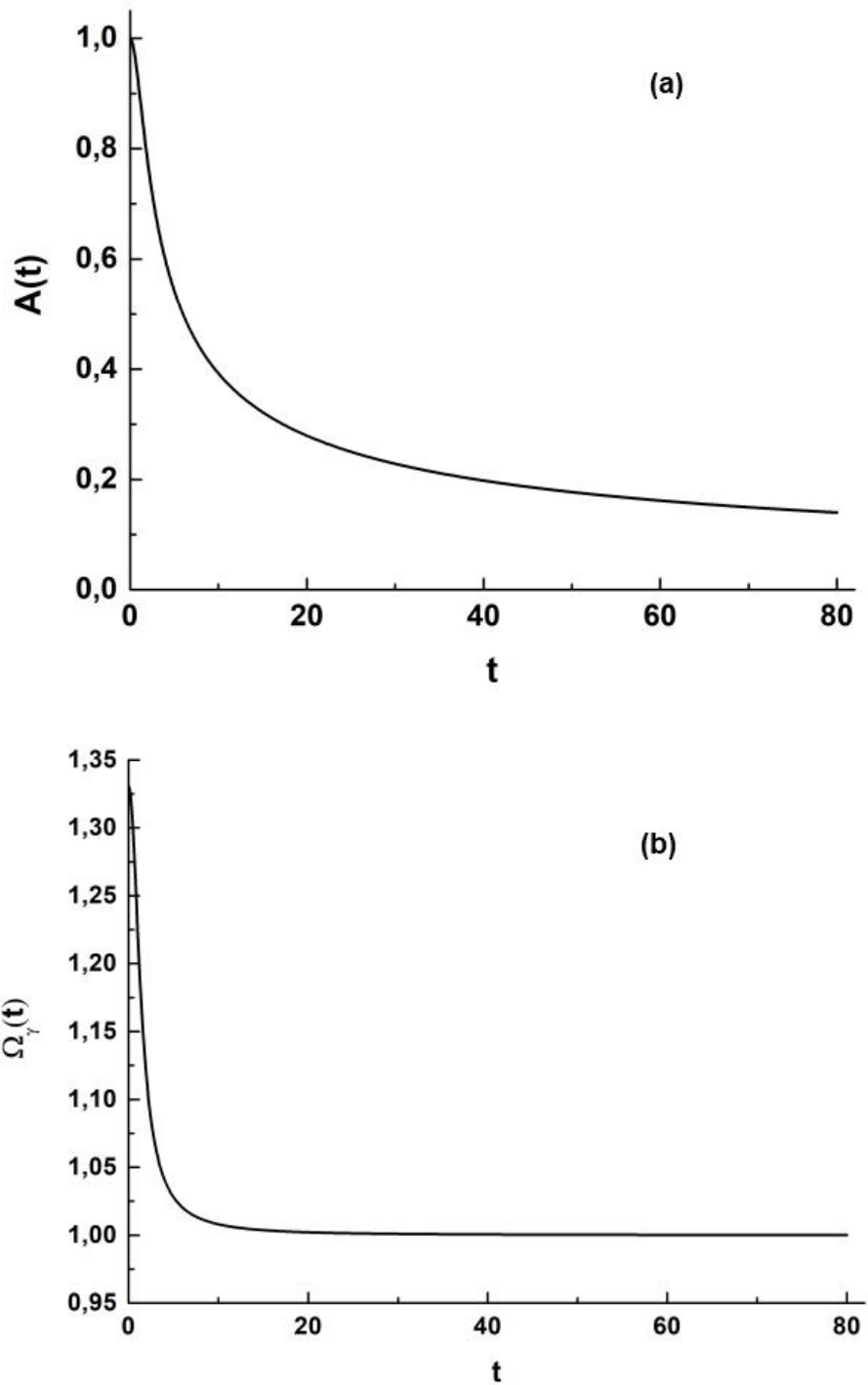

*Fig.3* The time dependencies of the amplitude (a) and the derivative of the full phase (b) of the addition to the reverse kink length in the quasimode regime.

The kink slope can vary not only through altering the effective kink length $l(t)$, but it also steepens through changing another collective variable, which can be introduced as follows:

$$u(x,t) = 2\{\arctan\exp(x + i\sqrt{\rho(t)}) + \arctan\exp(x - i\sqrt{\rho(t)})\}. \tag{20}$$

Naturally, expression (20) is a real function and can be written as

$$u(x,t) = \pi + 2\arctan\left(\frac{\sinh x}{\cos\sqrt{\rho(t)}}\right). \tag{21}$$

When parameter $\rho$ is constant and small $\rho \ll 1$, then we obtain the kink with the addition to its form

$$u(x,0) = 4\arctan\exp(x) + \frac{\rho \sinh x}{\cosh^2 x} \qquad (22)$$

Comparing the addition in (22) with the first term of the expression (7), we see that they coincide as functions and differ only by constant factors.

There is once more possibility to deform the exact kink. The following ansatz can be used to describe the dilation of the kink

$$u(x,t) = 2\{\arctan\exp(x + \sqrt{r(t)}) + \arctan\exp(x - \sqrt{r(t)})\}, \qquad (23)$$

or

$$u(x,t) = \pi + 2\arctan\left(\frac{\sinh x}{\cosh\sqrt{r(t)}}\right), \qquad (24)$$

where the collective variable $R(t) = \sqrt{r(t)}$ is a half of a distance between two $\pi$-subkinks. When $r$ is constant and small $r \ll 1$ we again find the same function (22) for the addition to the kink shape but with a coefficient of the opposite sign:

$$u(x,t) = 4\arctan\exp(x) - \frac{r \sinh x}{\cosh^2 x} \qquad (25)$$

Thus a transition from the real $r$ over zero to the imaginary $i\rho$ in the expressions (22–24) means the transition from a contraction of the separation between subkinks to the kink slope steepening.

Now we consider the Cauchy problem of Eq. (14) for the addition $\varphi(x,t)$ to the exact kink (4) with initial conditions: $\varphi(x,0) = \rho \sinh x / \cosh^2 x$ and $\varphi_t(x,0) = 0$. In the representation $\varphi_2(x,t) = \int_{-\infty}^{\infty} D_k \psi_k(x) \cos(\omega(k)t) dk$ the coefficients $D_k$ are defined from the initial conditions as follows

$$D_k = \sqrt{\frac{\pi}{8}} \frac{\omega(k)}{\cosh\frac{\pi}{2}k} \qquad (26)$$

and after substitution we find the addition function in the form

$$\varphi_2(x,t) = \frac{\rho}{4} \int_{-\infty}^{\infty} \frac{\cos(\sqrt{1+k^2}\,t)}{\cosh\frac{\pi}{2}k} (\tanh x \cos kx + k \sin kx) dk. \qquad (27)$$

It appears that main features of evolution of the additions $\varphi_1(x,t)$ and $\varphi_2(x,t)$ are similar because the integral in Eq. (27) determines the main contribution to the integral in the expression (16). Therefore, we call such an oscillating kink behavior as the second quasimode regime. We show the pictures of the evolution of the addition $\varphi_2(x,t)$ in Fig. 4 over the interval $\Delta t$ less than the period $T_0 = 2\pi$, namely $\Delta T = 8T_0/9$, in order to show the oscillation of the kink slope evidently and, as a result, the generation of the radiation taking away its energy.

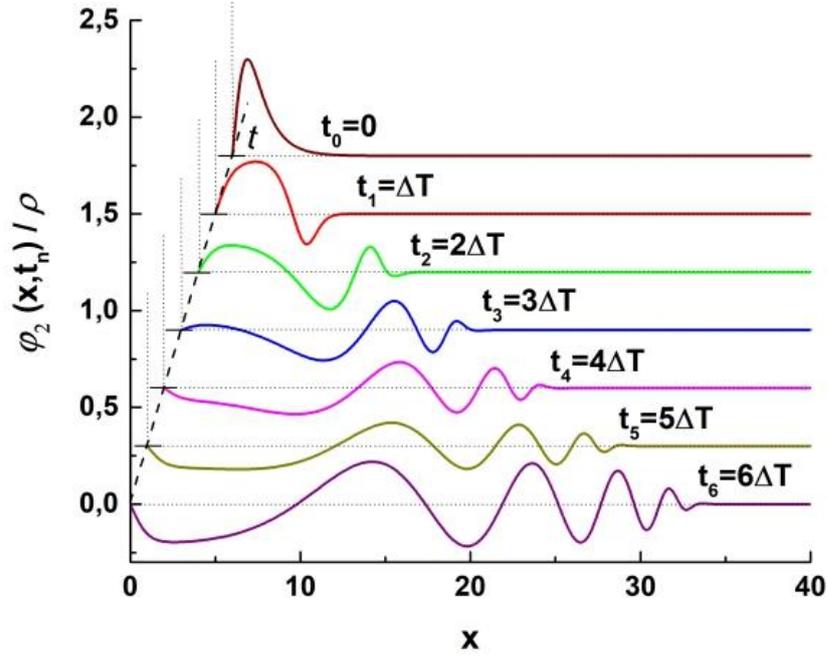

*Fig.4* The evolution of the addition to the SG kink in the second quasimode regime.

Proceeding similarly to the previous calculations, we find the expression for the addition to the reverse kink length

$$\Delta\kappa_2(t) = \frac{1}{2}\frac{\partial\varphi_2(x,t)}{\partial x}\bigg|_{x=0} = \frac{\rho}{4}\int_0^\infty \frac{(1+k^2)\cos\left(\sqrt{1+k^2}\,t\right)}{\cosh\frac{\pi}{2}k}dk. \qquad (28)$$

Its oscillating time dependence is shown in Fig. 5 and it is qualitatively similar to the time dependence $\Delta\kappa_1(t)$ (Fig. 2).

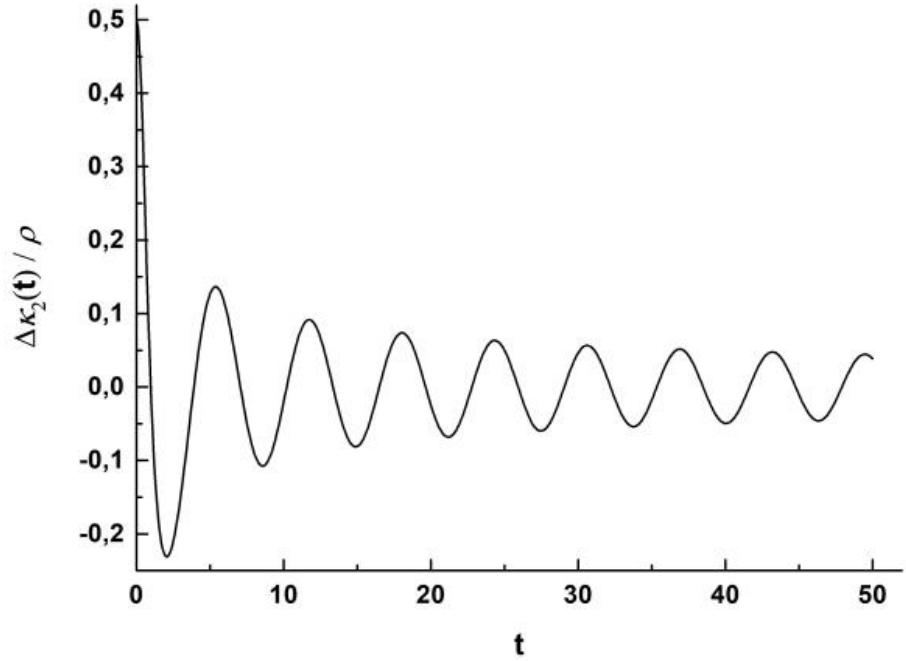

*Fig.5* Weakly damped oscillations of the addition to the reverse kink length in the second quasimode regime.

As we see from the solution (9) the definite linear combination of $\varphi_1(x,t)$ and $\varphi_2(x,t)$ solves the problem of transformation of the SG kink to the RGSE kink with the addition (7). Now we concentrate on the difference between $\varphi_1(x,t)$ and $\varphi_2(x,t)$. We show how it can be used to excite the kink that can throw the wave packets. We choose the static initial kink profile in the form:

$$u(x,0) = 2\{\arctan \exp(\kappa_* x + i\sqrt{\rho_*}) + \arctan \exp(\kappa_* x - i\sqrt{\rho_*})\}, \qquad (29)$$

where constant $\kappa_* = 1-\eta$ and $\rho_* = 2\sqrt{\eta}$, and parameter $\eta \ll 1$. Then in the first approximation with respect to parameter $\eta$ the initial addition to the exact kink (4) becomes as follows

$$\varphi(x,0) = 2\eta\left(2\frac{\sinh x}{\cosh^2 x} - \frac{x}{\cosh x}\right) \qquad (30)$$

Notice that the same function with the opposite sign can be obtained using a combination of the kink form (12) and the ansatz (23) after the following choice of the constant parameters $\kappa = 1+\eta$ and $r = 2\sqrt{\eta}$, where $\eta$ is small as before.

The solution of the Cauchy problem of Eq. (14) in the case of the initial condition (30) and $\varphi_t(x,0) = 0$ looks like the following:

$$\varphi_3(x,t) = \eta \int_{-\infty}^{\infty} \frac{k^2 \cos(\sqrt{1+k^2}\,t)}{(1+k^2)\cosh\frac{\pi}{2}k}(\tanh z \cos kz + k \sin kz)dk \qquad (31)$$

The evolution of the addition to the exact kink profile is presented in Fig. 6. We call this oscillating kink behavior as the damped quasimode regime. As can be seen, the kink gives out almost all its energy to the emitted wave packet that goes quite local. Thus in order to use the kink as a source of the wave packets in the SG system we can prepare the initial deformed kink by its simultaneous smoothing and steepening through the special choice of the parameters $\kappa$ and $\rho$ respectively.

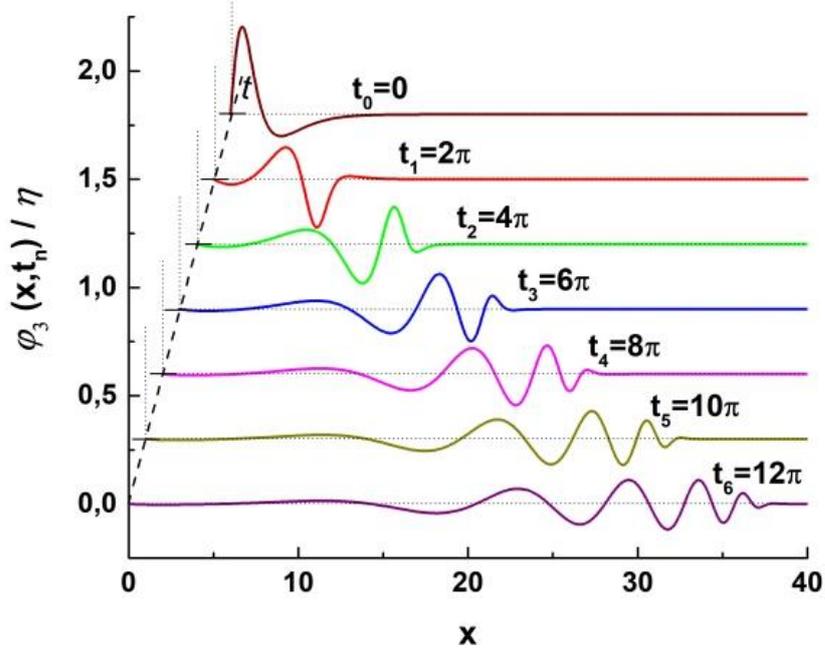

*Fig.6* The evolution of the addition to the SG kink with a formation of the wave packet in the damped quasimode regime.

The expression for the addition to the reverse kink length is found following the previous schemes:

$$\Delta\kappa_3(t) = \frac{1}{2}\frac{\partial\varphi_3(x,t)}{\partial x}\bigg|_{x=0} = \eta\int_0^\infty \frac{k^2 \cos\left(\sqrt{1+k^2}\,t\right)}{\cosh\frac{\pi}{2}k}dk. \tag{32}$$

Its time dependence is shown in Fig. 7. The fast Fourier transform of the dependence gives a maximum of amplitude at the frequency $\Omega_2 = 1.1304$, which is certainly above the lowest frequency edge $\Omega_0 = 1$ and distinctly close to Rice's frequency $\Omega_R = 2\sqrt{3}/\pi = 1.1027$ [7]. Notice that the period of the oscillation with the frequency $\Omega_2$ is very close to the time interval $\Delta T$ which separates configurations in Fig. 4. The addition to the reverse kink length can be represented again in the form $\Delta\kappa_3(t) = \eta B(t)\cos(t + \delta(t))$, where the amplitude and the phase are given by $B(t) = |I(t)|$ and $\delta(t) = \text{Arg}\,I(t)$ respectively, and the integral $I(t)$ is defined as

$$I(t) = \int_0^\infty \frac{k^2 \exp\left(i\left(\sqrt{1+k^2}-1\right)t\right)}{\cosh\frac{\pi}{2}k}dk. \tag{33}$$

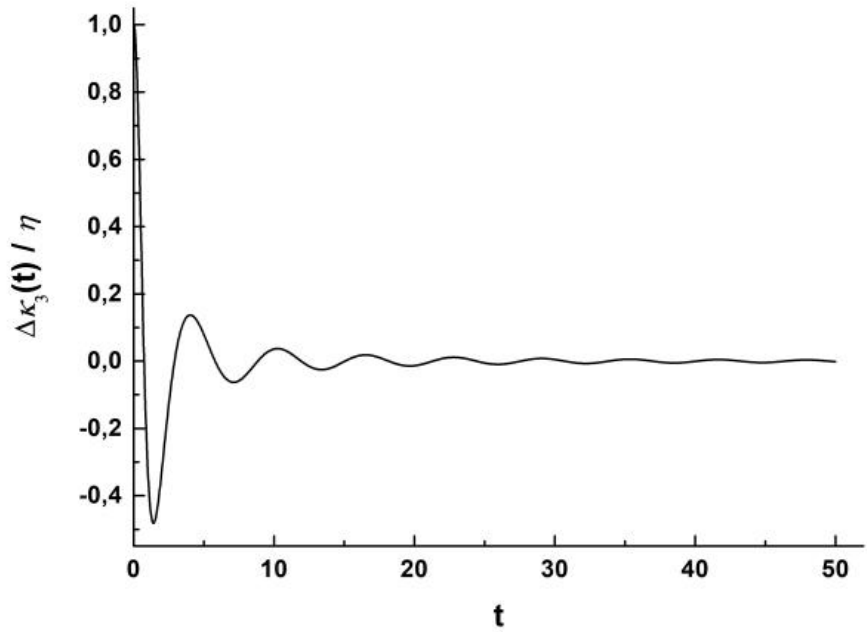

*Fig.7* Fast decaying oscillations of the addition to the reverse kink length in the damped quasimode regime.

The time dependencies for $B(t)$ and $\delta(t)$ are calculated numerically as well as the time derivative of the full phase $\Omega_\delta(t) = 1 + \dfrac{d}{dt}\delta(t)$. Corresponding curves for $B(t)$ and $\Omega_\delta(t)$ are shown in Fig.8 a,b. They are monotonic decreasing functions, which go to their asymptotics evidently faster than analogous dependencies of the addition $\Delta\kappa_1(t)$ to the reverse kink length.

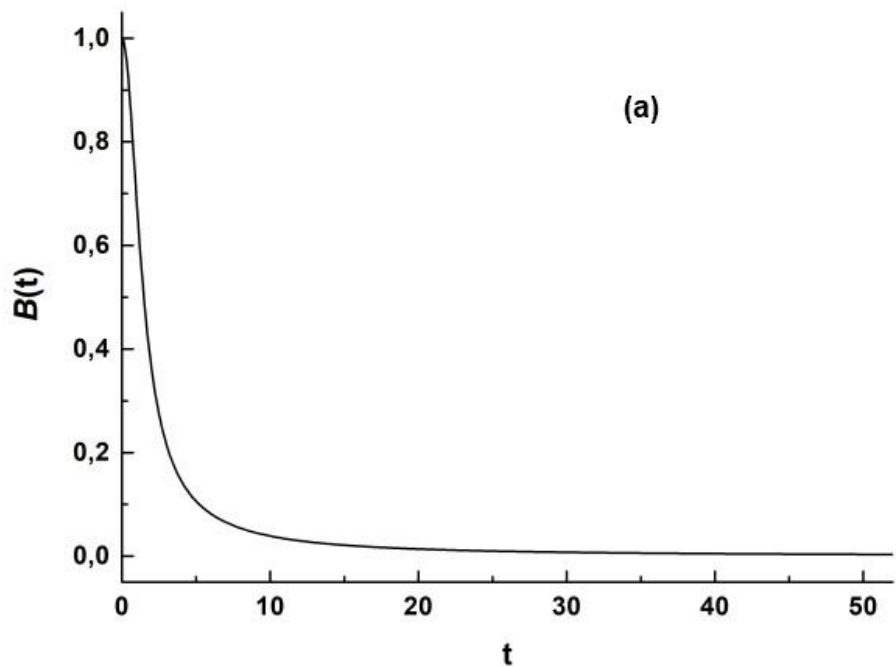

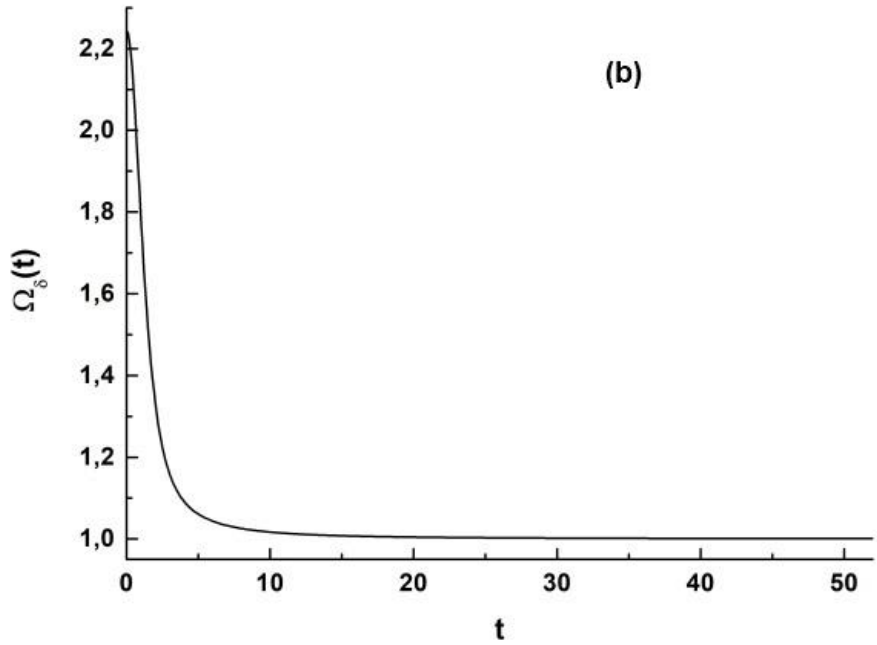

*Fig.8* The time dependencies of the amplitude (a) and the derivative of the phase (b) of the addition to the reverse kink length in the damped quasimode regime.

Thus all types of internal kink oscillations after a transient period of evolution show a quasi-local character of motion with slowly decaying frequencies just above the lowest edge $\Omega_0 = 1$. They can be considered as *lagrangian collective coordinates* describing periodic changing of the slope and the effective length of kink profiles. Further inclusion of dissipation in the consideration is absolutely necessary [8], because of loss of the kink energy, which is expended to excitation of linear waves, and external periodic forces can compensate for this energy leakage. The question about the energy radiation by the nonequilibrium kink in the framework of the SG equation (2), including a generation of breathers, we discuss in the next section.

## Breather generation and energy radiation by nonequilibrium kink

The analysis of small oscillations of a nonequilibrium kink by means of the perturbation theory led to the linearized equation and finally to the solution of the quasimode problem. The general problem on the evolution of the nonequilibrium kink (12) with considerable deviation from the exact solution (4) has to be solved in the framework of the nonlinear SG equation (2). In order to solve the Cauchy problem of the integrable sine-Gordon equation, the inverse scattering method is usually used [1]. In point of fact, it is enough to solve the direct scattering problem associated with the sine-Gordon equation. This task for the kink (4) was formulated in Ref. 25 in the general form, in particular, the exact profile (4) was considered to have the non-zero velocity $u_t(x,0) = 2\mu/\cosh(x)$ with an arbitrary constant parameter $\mu$. As a result, authors of [25] achieved success in the construction of a sequence of exact multibreather wobbling kinks which correspond to the reflectionless spectral problem.

We use the direct scattering problem formulated in [25] to analyze the Cauchy problem for initially static kink $u(x,0) = 4\arctan(\exp(\kappa x))$ with an arbitrary value of the parameter $\kappa$. We reduce the task to solving the spectral problem of the well-known one-dimensional Schrödinger operator, which eigenvalues determine the number of breathers generated from the nonequilibrium kink and values of their parameters. The final solution presents the multibreather wobbling kink, in which all breathers are located strictly in the central part, and the excess energy is emitted as linear wave packets. Eventually, the radiation energy can be found as a function of the effective length of the initial kink profile.

It is known [1], that the spectral problem associated with the SG equation in the inverse scattering method can be formulated as follows:

$$\bar{\mathbf{J}}_\xi = \frac{i\lambda}{2}U_1\bar{\mathbf{J}} + \frac{1}{8i\lambda}U_2\bar{\mathbf{J}} + \frac{i}{4}U_3\bar{\mathbf{J}}, \tag{34}$$

where $\lambda$ is the spectral parameter, $\bar{\mathbf{J}} = \begin{pmatrix} \psi_1 \\ \psi_2 \end{pmatrix}$ is the Jost functions and matrices are equal to:

$$U_1 = \begin{pmatrix} 1 & 0 \\ 0 & -1 \end{pmatrix}, \quad U_2 = \begin{pmatrix} \cos u & -i\sin u \\ i\sin u & -\cos u \end{pmatrix}, \quad U_3 = \begin{pmatrix} 0 & u_x - u_t \\ u_x - u_t & 0 \end{pmatrix} \tag{35}.$$

In [25], the matrix eigenvalue problem was significantly simplified. After carrying out the following replacements

$$z = 2\lambda, \qquad S_\pm = \frac{1}{4}\left(z \pm \frac{1}{z}\right), \tag{36},$$

the problem was rewritten as

$$4i\bar{\mathbf{J}}_x = \begin{bmatrix} 4S_- + z^{-1}(1-\cos(u)) & -z^{-1}\sin(u) - i(u_x + u_t) \\ -z^{-1}\sin(u) + i(u_x + u_t) & -4S_- - z^{-1}(1-\cos(u)) \end{bmatrix} \bar{\mathbf{J}} \tag{37}$$

and by the use of the rotation transformation

$$\mathbf{J}(x;z,t) = A\bar{\mathbf{J}} = \begin{bmatrix} \cos\left(\frac{u}{4}\right) & \sin\left(\frac{u}{4}\right) \\ -\sin\left(\frac{u}{4}\right) & \cos\left(\frac{u}{4}\right) \end{bmatrix} \bar{\mathbf{J}} \tag{38}$$

it was reduced to the form:

$$i\mathbf{J}_x = \begin{bmatrix} S_- \cos\left(\dfrac{u}{2}\right) & -S_+ \sin\left(\dfrac{u}{2}\right) - \dfrac{1}{4} i u_t \\ -S_+ \sin\left(\dfrac{u}{2}\right) + \dfrac{1}{4} i u_t & -S_- \cos\left(\dfrac{u}{2}\right) \end{bmatrix} \mathbf{J}. \qquad (39)$$

As can be seen, for the class of initial conditions with the time derivative $u_t(x,0)=0$, the direct scattering problem has an even simpler form:

$$i\mathbf{J}_x = \begin{bmatrix} S_- \cos\left(\dfrac{u}{2}\right) & -S_+ \sin\left(\dfrac{u}{2}\right) \\ -S_+ \sin\left(\dfrac{u}{2}\right) & -S_- \cos\left(\dfrac{u}{2}\right) \end{bmatrix} \mathbf{J}. \qquad (40)$$

Starting from Eq. (40), we analyze the problem of the evolution of the initially static nonequilibrium kink profile $u(x,0)=4\arctan(\exp(\kappa x))$. It is known that the spectral eigenvalues lie exactly on the imaginary axis of the upper half-plane of a complex parameter $z$ for kinks and on the circle of the unity radius for breathers, respectively. Interesting in the kink and breather solutions, we introduce for the spectral parameter the notation $z=\exp(i\chi)$ that is valid for unmovable breathers and also for the exact kink (4) when $\chi=0$ and $z=1$. For this choice of $z$ parameters $S_-$ and $S_+$ become

$$S_- = \dfrac{i}{2}\sin\chi, \qquad S_+ = \dfrac{1}{2}\cos\chi \qquad (41)$$

Introducing notations $U_1 = \dfrac{1}{2}\sin\chi\cos\dfrac{u}{2}$ and $U_2 = \dfrac{1}{2}\cos\chi\sin\dfrac{u}{2}$, we rewrite the matrix equation as the system

$$\dfrac{d\psi_1}{dx} = U_1 \psi_1 + i U_2 \psi_2, \qquad \dfrac{d\psi_2}{dx} = i U_2 \psi_1 - U_1 \psi_2. \qquad (42)$$

We define conjugate operators

$$L^- = \dfrac{d}{dx} + U_1, \qquad L^+ = -\dfrac{d}{dx} + U_1 \qquad (43)$$

and obtain the following compact form the system (42)

$$L^+ \psi_1 = -i U_2 \psi_2, \qquad L^- \psi_2 = i U_2 \psi_1. \qquad (44)$$

Then the equations for real and imaginary parts of functions $\psi_{1,2} = g_{1,2} + i v_{1,2}$ take the form

$$L^+ g_1 = U_2 v_2, \qquad L^- v_2 = U_2 g_1; \qquad (45)$$

$$L^+ v_1 = -U_2 g_2, \qquad L^- g_2 = -U_2 v_1. \tag{46}$$

By replacing $g_1 \to v_1$ and $v_2 \to -g_2$, it is easy to see that these systems are equivalent. We aim to obtain the differential equation of the Schrödinger-like type from the system (45). Therefore we apply the operator $L^-$ to the first equation from the system (45)

$$\left( L^- L^+ - U_2^2 - \frac{d \ln U_2}{dx} L^+ \right) g_1 = 0 \tag{47}$$

and similarly the operator $L^+$ to the second equation (46) of this system

$$\left( L^+ L^- - U_2^2 + \frac{d \ln U_2}{dx} L^- \right) v_2 = 0, \tag{48}$$

and find the following explicit form of the second order differential equation:

$$\left( -\frac{d^2}{dx^2} \pm \left( \frac{dU_1}{dx} - U_1 \frac{d \ln U_2}{dx} \right) + U_1^2 - U_2^2 + \frac{d \ln U_2}{dx} \frac{d}{dx} \right) \binom{g_1}{v_2} = 0. \tag{49}$$

The upper sign refers to the function $g_1$ and the lower sign to the function $v_2$, respectively. In order to exclude the first derivate, we perform the substitution, for definiteness, for function $g_1 = \sqrt{U_2} f$ and obtain the equation of the spectral problem of the Schrödinger operator

$$\left( -\frac{d^2}{dx^2} + W(x, \sin \chi) \right) f = 0, \tag{50}$$

where the potential well is as follows

$$W(x, \sin \chi) = \frac{1}{4} \left( \frac{d \ln U_2}{dx} \right)^2 - \frac{1}{2} \frac{d^2 \ln U_2}{dx^2} + U_1^2 - U_2^2 + U_1 \frac{d}{dx} \ln \frac{U_1}{U_2}. \tag{51}$$

This potential well can be expressed explicitly in the terms of the function $u$

$$W(x, \sin \chi) = \frac{1}{4} \left( \left( \frac{d \ln \sin \frac{u}{2}}{dx} \right)^2 - 2 \frac{d^2 \ln \sin \frac{u}{2}}{dx^2} + \sin^2 \chi - \sin^2 \frac{u}{2} - 2 \sin \chi \cos \frac{u}{2} \frac{d \ln \tan \frac{u}{2}}{dx} \right) \tag{52}$$

The equation (50) with the potential (52) can be used for any static initial condition $u(x,0)$ to determine easily whether or not breathers are generated and to find their parameters using developed methods of analysis and solving the one-dimensional Schrödinger equation [24].

In the case of the initial condition in the form of a nonequilibrium kink (12) the functions in the potential (52) are as follows:

$$\sin\frac{u}{2} = \frac{1}{\cosh\kappa x}, \qquad \cos\frac{u}{2} = -\tanh\kappa x \qquad (53)$$

As a result, after the substitution (53) into Eqs. (50) and (52) and the introduction of a variable $y = \kappa x$ we obtain the following eigenvalue equation:

$$\left\{-\frac{d^2}{dy^2} - \frac{1}{4}\left(\frac{1}{\kappa^2} - 1\right)\frac{1}{\cosh^2 y}\right\}f = -\frac{1}{4}\left(1 - \frac{\sin\chi}{\kappa}\right)^2 f. \qquad (54)$$

Using the formula for discrete levels of this well-known equation from [24]:

$$\Lambda_n = -\frac{1}{4}\left(-(1+2n) + \sqrt{1+4U_0}\right)^2, \qquad (55)$$

where $\Lambda_n = -\frac{1}{4}\left(1 - \frac{\sin\chi}{\kappa}\right)^2$ and $U_0 = \frac{1}{4}\left(\frac{1}{\kappa^2} - 1\right)$ we obtain the condition on the number and parameters of breathers arising from the unequilibrium kink (12):

$$\varepsilon_n \equiv \sin\chi_n = 1 - 2n\kappa. \qquad (56)$$

It is easy to see that at $\kappa = 1$, the number of breathers is $n = 0$. When $\kappa = \frac{1}{2}$ and $n = 1$, the first breather appears, and when $\kappa_n = \frac{1}{2n}$ then $n$-th breather arises.

In general, the solution presents the multi-frequency oscillating kink that generates linear waves of the continuous spectrum. This complex kink is a nonlinear superposition of the kink and breathers with parameters $\varepsilon_n$ which determine their frequencies $\omega_n = \sqrt{1-\varepsilon_n^2} = \cos\chi_n$ and eventually their energies. In the integrable SG system the integral of the energy

$$E = \int_{-\infty}^{\infty}\left(\frac{1}{2}\left(u_t^2 + u_x^2\right) + 1 - \cos u\right)dx \qquad (57)$$

is the additive function of energies of all the nonlinear and linear excitations [1]. Therefore, the energy of the $n$-th breather, $E_n = 16\varepsilon_n$, is conserved and does not depend on its interaction with other excitations. Returning to the effective length $l = 1/\kappa$ of the nonequilibrium kink, we obtain the expression for the energy of the $n$-th emerging breather $E_n(l) = 16\left(1 - \dfrac{2n}{l}\right)$. Now it is easy to calculate the energy of radiation arising during the evolution of a nonequilibrium kink:

$$E_r(l) = E_K(l) - E_{brs}(l) - E_K^0,  \qquad (58)$$

where the energy of the initial profile of the kink is $E_K(l) = 4\left(l + \dfrac{1}{l}\right)$, the energy of all breathers $E_{brs}(l) = 16\sum_{n=1}^{N}\left(1 - \dfrac{2n}{l}\right)$, where their total number is equal to the integer part of $l/2$, i.e. $N = \left[\dfrac{l}{2}\right]$, and the energy of the static kink as a final state of its evolution is $E_K^0 = 8$. After calculating we obtain for the energy of all breathers

$$E_{brs}(l) = 16\sum_{n=1}^{N}\left(1 - \dfrac{2n}{l}\right) = 16\left[\dfrac{l}{2}\right]\left(1 - \dfrac{\left[\dfrac{l}{2}\right] + 1}{l}\right), \qquad (59)$$

and get finally the following dependence:

$$E_r(l) = 4\left(l + \dfrac{1}{l}\right) - 16\left[\dfrac{l}{2}\right]\left(1 - \dfrac{\left[\dfrac{l}{2}\right] + 1}{l}\right) - 8. \qquad (60)$$

This function is shown in Fig. 9.

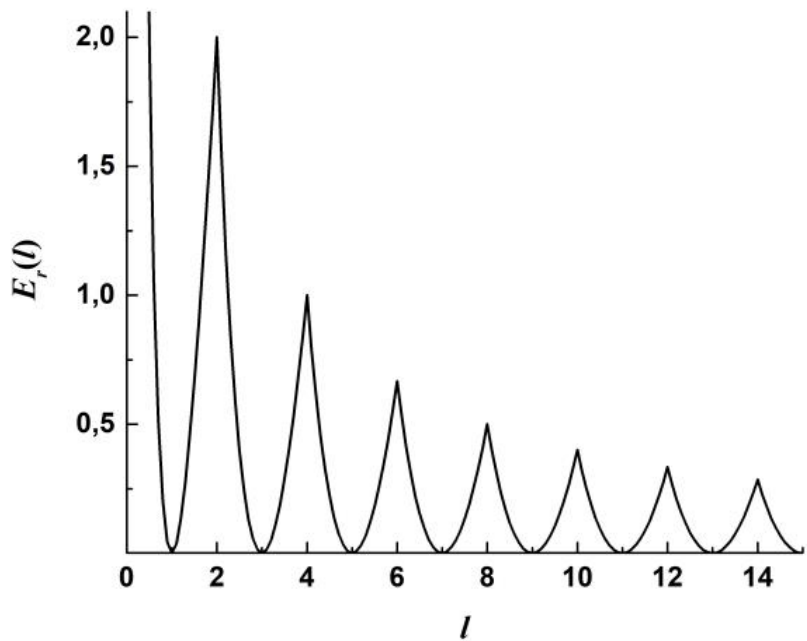

*Fig.9* The dependence of the radiation energy on the effective length of the initial kink profile.

It is easy to verify that for an even integer $l = 2m$, at the moment of birth of the next breather, the radiation energy reaches its maximum $E_r(2m) = \dfrac{2}{m}$. For odd numbers $l = 2m+1$, the radiation energy is absent, $E_r(2m+1) = 0$, and all the energy of the system is localized in space in the form of an ensemble of breathers settling on the kink, i.e. in the form of a multi-frequency oscillating kink. In particular, in the case $l = 3$ and $n = 1$ the exact wobbling kink is formed from the unequilibrium profile (12). The wobbling kink presents a nonlinear superposition of the kink (12) with $\kappa = 1/3$ and the breather with parameters $\varepsilon_1 = 1/3$ and the frequency $\omega_1 = 2\sqrt{2}/3$ as it follows from Eq. (56). Notice that the explicit cosine form of this solution was found first in the Ref. 25.

No breather exists up to the length value $l = 2$. In the vicinity of the point $l = 1$, there is the single oscillating kink which relaxes to the exact solution (4) in the quasimode regime described in the previous section. Therefore no internal oscillation appears with the frequency below $\Omega_0 = 1$ from the slightly perturbed kink (12) at the nonlinear evolution stage.

Analysis of the Cauchy problems with initial conditions (22) and (29) in the previous section dealt with linear waves around the kink and indicated only the formation of wave packets. In order to clarify whether breather birth happens at the nonlinear stage, we have to solve the equations of the direct scattering problem. In this case, the obtained equation (50) with the potential (52) has a certain advantage. The substitution of the exact kink solution (4) into the potential (52) leads to the absence of the potential well in the Eq. (54) at all. Therefore, any small addition to the shape of this kink is entirely responsible for the creation of the own potential and its specific features. After the substitution of $u = u_K + \varphi$ and linearization of $W(u, \sin \chi)$ with respect to small $\varphi$, we find new potential $V(\varphi, u_K, \sin \chi)$ for the Schrödinger equation (50), the discrete levels of which determine a presence and parameters of the breather. In the case of the initial conditions (12) with the small parameter $\eta$, the corresponding equation is found directly from Eq. (54) after the substitution $\kappa = 1 + \eta$ and the linearization with respect to $\eta$. In accordance with the above conclusion, the linearized equation has no discrete eigenvalue. Further, we performed calculations for the initial condition (22) and found the following equation

$$\left\{ -\frac{d^2}{dx^2} - \frac{\rho}{4}\left( \frac{1+\sin \chi}{\cosh^2 x} - \frac{3}{\cosh^4 x} \right) \right\} f = -\frac{1}{4}(1 - \sin \chi)^2 f, \qquad (61)$$

in which the suitable discrete eigenvalue is also absent in the case of small parameters $\rho$ and $\chi$.

However, it does not mean that the arbitrary small static deformation of the exact kink (4) can not cause the breather birth. For example, the following initial condition for the addition to the kink shape with a free parameter $\varepsilon$

$$\varphi(x,0) = -4\arctan\left(\frac{\varepsilon\left(\dfrac{\tanh \varepsilon x}{\cosh x} + \dfrac{\tanh x}{\cosh \varepsilon x}\right)}{1 - \varepsilon\left(\tanh x \tanh \varepsilon x - \dfrac{1}{\cosh x \cosh \varepsilon x}\right)}\right) \qquad (62)$$

generates the exact wobbling kink solution of the SG equation. We present here this wobbling kink in the form that shows evidently that the breather, settling on the exact kink (4), can possess the small amplitude $\varepsilon$ and the frequency $\omega = \sqrt{1-\varepsilon^2}$ that is close to unity:

$$u_{WK} = \pi + 4\arctan\left(\frac{\tanh\dfrac{x}{2} - \varepsilon\left(\tanh \varepsilon x + \tanh\dfrac{x}{2} \cdot \dfrac{\cos \omega t}{\cosh \varepsilon x}\right)}{1 - \varepsilon\left(\tanh\dfrac{x}{2} \cdot \tanh \varepsilon x - \dfrac{\cos \omega t}{\cosh \varepsilon x}\right)}\right) \qquad (63)$$

Finally notice, while the breather standing alone is evidently the even function [26], the small amplitude breather on the kink background is the odd function and has the form of the kink internal mode.

## Conclusion

The main findings of this study are as follows:

1. Nonstationary dynamics of topological defects and inhomogeneities described by the sine-Gordon equation is studied. The evolution of nonequilibrium profiles of these topological objects is considered, and their oscillation regimes of the approach to static configurations are investigated in terms of nonlinear excitations of the sine-Gordon equation, kinks and breathers. In order to describe explicitly oscillatory behavior of these objects, the Cauchy problem for the equation linearized near the exact static kink is solved for the small addition to the kink shape. The obtained solution describes explicitly the regime of long-living oscillations of the kink, named in Ref. 5 the quasimode of kink. It is shown that the kink deformations leading to its initial compression and tension evolve as weakly damped oscillations with the frequency lying just above the lowest edge of the continuous spectrum. The specially combined deformation leads to a great rate of damping the kink oscillation and to the formation of the well-defined wave packet carrying away all the exceed energy of the initial kink profile. The patterns of the kink evolution and the time dependencies of its effective length are presented.

2. The nonlinear stage of evolution of the kink with the deformed slope is investigated in the framework of the inverse scattering method. In fact, the direct scattering problem solution appears to be enough to find exactly the conditions of arising the breathers from the nonequilibrium kink and to determine their amplitudes, frequencies and finally their energies. In general, we reduce the direct scattering problem, considered in Ref. 25, first to our knowledge, to solving the spectral problem of the one-dimensional Schrödinger equation. In the case of the nonequilibrium kink, this equation corresponds to the well-known one in quantum mechanics [24] with the famous spectrum of discrete levels. After finding parameters of breathers for the multi-frequency oscillating kink, we are able to consider the complex solution consisting of namely the kink, arising breathers and wave packets generated by the kink. We calculate the dependence of the radiation energy on the effective length of the initial kink profile and analyze the structure of the ensemble of linear and nonlinear excitations depending on this parameter. At last, we note the advantage of the obtained Schrödinger equation that using the equation we are able to determine whether or not a small deformation of the exact kink (4) leads to the small-amplitude breather birth.

ACKNOWLEDGMENTS